\documentclass{appolb}
\usepackage{epsfig}
\def\pp{p-p~}
\def\sr17{$\sqrt{s}$~=~17~GeV~}

\def\prc{{\em Phys. Rev. {\bf C} }}
\def\prd{{\em Phys. Rev. {\bf D} }}

\def\epjc{{\em Eur. Phys. J. {\bf C} }}

\def\plb{{\em Phys. Lett. {\bf B} }}

\newcommand{\pbar}{$\rm\overline{p}$}

\newcommand{\be}{\begin{equation}}
\newcommand{\ee}{\end{equation}}

\def\muB{$\mu_B$}

\begin{document}
\pagestyle{plain}
\newcount\eLiNe\eLiNe=\inputlineno\advance\eLiNe by -1
\title{The Thermal Model at the Large Hadron Collider} 
\author{J.~Cleymans
\address{UCT-CERN Research Centre and Department of Physics,\\ University
of Cape Town, South Africa}}
\maketitle
\abstract{
A discussion is presented of results 
with identified particles at the Large Hadron Collider.
Possible deviations from the 
standard statistical distributions are investigated by considering 
in detail results obtained using the Tsallis distribution.  
Matter-antimatter production is discussed within the framework
of chemical equilibrium in \pp and heavy ion collisions.
}
%
\section{The Hadronic World}
The available energy  for heavy ions at the Large Hadron Collider (LHC) is
$\sqrt{s} = 2760$ AGeV yet the  observed temperature is only of the
order of $T \approx $ 0.160 GeV.  To understand this enormous change 
from the initial state to the final state we 
first clarify how  this temperature is determined.  
There are several  independent ways of doing this.
\begin{enumerate}
\item From the number of hadronic resonances listed in the particle
data booklet~\cite{PDG}. This method was first proposed by
Hagedorn~\cite{hagedorn}. Note that this involves no 
transverse momentum spectrum, no energy distribution, only the number
of particles listed in the PDG~\cite{PDG} table.  
A recent updated version of this determination is shown in 
Fig.~\ref{worku} which shows the logarithm of the number of resonances 
below a certain mass~\cite{sound_worku}.  The fitted line corresponds to a Hagedorn 
temperature of
\begin{equation}
T_H = 174 \pm 11~~~\mathrm{MeV}.
\end{equation}
Other recent determinations are consistent with this 
value~\cite{india,poland1,poland2,poland3}.
At  masses above 3 GeV  the increase stops  due
to the difficulty in identifying heavy hadronic resonances, a situation which will probably not be resolved experimentally over the next years.
\begin{figure}

\begin{center}
\includegraphics[width=0.6\linewidth,height=6cm]{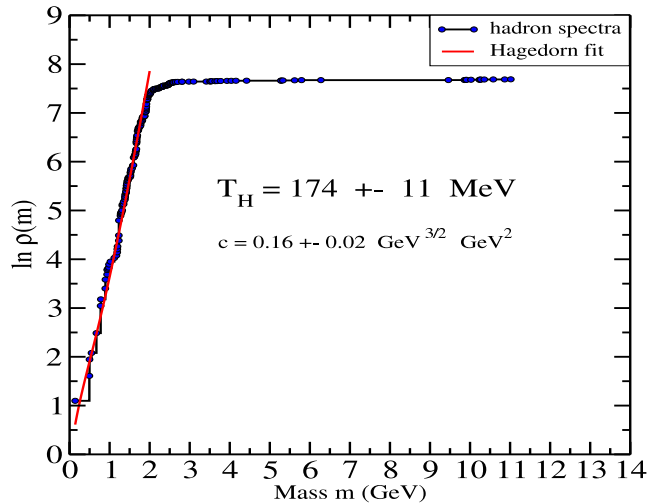}
\label{worku}
\caption{Cumulative number of hadronic resonances as a 
function of $m$~\cite{sound_worku}.
The hadronic data include baryons, mesons and also heavy resonances
made up of charm and bottom quarks.}
\end{center}
\end{figure}
\item The  multiplicity of particles in the final state.  This has been
an ongoing effort for the past two decades \cite{stachel,becattini,comparison}.  
Again this involves no transverse momentum or energy distribution. In this case it is only
the number of identified particles in the final state.
The temperature at $\mu_B=0$ is remarkably close to the original 
Hagedorn temperature~\cite{hagedorn} obtained by 
summing the number of hadronic resonances. 
\item The critical temperature determined from Lattice QCD
is again remarkably close to the Hagedorn temperature and the 
chemical freeze-out temperature at $\mu_B = 0$~\cite{fodor,Cheng:2009zi}.
\item The temperature can also be determined from the slope of the 
transverse momentum spectrum. This leads to a lower temperature, at least in
\pp collisions and will be discussed in detail below.
\end{enumerate}

\section{\label{transverse}Transverse Momentum Distribution}
An unusual form of the Tsallis distribution (sometimes
referred to as Levy-Tsallis) 
has gained prominence recently  
in high energy physics with high quality fits
of the transverse momentum distributions  made
by the STAR~\cite{Abelev:2006cs} and PHENIX~\cite{Adare} collaborations at the 
Relativistic Heavy Ion Collider 
and by the ALICE~\cite{Aamodt:2011zj} and CMS~\cite{Khachatryan:2011tm}
 collaborations at the Large Hadron Collider.

In the literature there exists more than one version of the Tsallis 
distribution~\cite{Tsallis:1987eu,Tsallis:1998ab}.
In this paper we investigate a version which 
 we consider suited for
describing results in high energy particle physics.
Our main guiding criterium will be thermodynamic consistency which
has not always been implemented 
correctly. 
The explicit form which we use  is~\cite{cleymans_worku}: 
\begin{equation}
  \frac{{\rm d}^2N}{{\rm d}p_{\rm T}~{\rm d}y} =
gV\frac{p_{\rm T}m_{\rm T}\cosh y}{(2\pi)^2}
\left[1+(q-1)\frac{m_T\cosh y -\mu}{T}\right]^{q/(1-q)},
\label{Tsallis-B}
\end{equation}
where $p_T$ and $m_T$ are the transverse momentum and mass respectively, $y$
is the rapidity, $T$ and $\mu$ are the temperature and the chemical potential,
$V$ is the volume, $g$ is the degeneracy factor.
%

\begin{figure}[tbh]
\begin{center}
\includegraphics[width=0.7\linewidth,height=6cm]{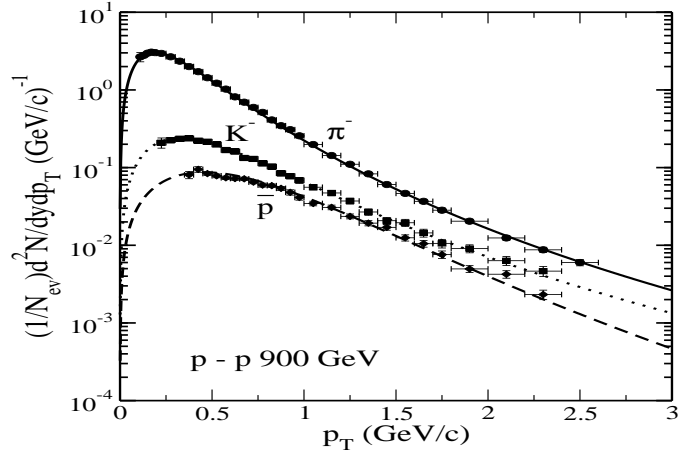}
\label{pikp}
\caption{Fit to the data~\cite{Aamodt:2011zj} using the Tsallis 
distribution~\cite{cleymans_worku}.}
\end{center}
\end{figure}

The motivation for preferring this form is presented in detail in the rest of this paper.
The parameterization given in Eq.~(\ref{Tsallis-B}) is close (but different) from
the one used by STAR, PHENIX, ALICE and CMS~\cite{Abelev:2006cs,Adare,Aamodt:2011zj, Khachatryan:2011tm}:
\begin{equation}
  \frac{{\rm d}^2N}{{\rm d}p_{\rm T}~{\rm d}y} = p_{\rm T} \frac{{\rm d}N}
  {{\rm d}y} \frac{(n-1)(n-2)}{nC(nC + m_{0} (n-2))} \left( 1 + \frac{m_{\rm T} - m_{0}}{nC} \right)^{-n},
\label{ALICE-CMS}
\end{equation}
where $n$, $C$ and $m_0$ are fit parameters.
The analytic expression used in Refs.~\cite{Abelev:2006cs,Adare,Aamodt:2011zj,Khachatryan:2011tm} 
corresponds to identifying 
\begin{equation}
n\rightarrow \frac{q}{q-1}
\end{equation}
and 
\begin{equation}
nC  \rightarrow \frac{T+m(q-1)}{q-1}.
\end{equation}
But differences do not allow for the above identification to be made 
complete due to an additional factor of the transverse mass on
the right-hand side and
a shift in the transverse mass.
They are close but not the same.
In particular, no clear pattern emerges for the values of $n$ and $C$ while
an interesting regularity is obtained for $q$ and $T$ as seen in 
Table 1.
\\
\begin{table}[b]
\begin{center}
\begin{tabular}{|c|c|c|}
\hline
Particle & $q$                &$T$ (GeV)      \\
\hline 
$\pi^+$  & 1.154  $\pm$0.036 & 0.0682 $\pm $0.0026 \\ 
$\pi^-$  & 1.146  $\pm$0.036 & 0.0704 $\pm$ 0.0027 \\
$K^+$    & 1.158  $\pm$0.142 & 0.0690 $\pm $0.0223  \\
$K^-$    & 1.157  $\pm$0.139 & 0.0681 $\pm$ 0.0217  \\
$K^0_S$  & 1.134  $\pm$0.079 & 0.0923 $\pm $0.0139  \\
$p$      & 1.107  $\pm$0.147 & 0.0730 $\pm$ 0.0425   \\
$\bar{p}$& 1.106  $\pm$0.158 & 0.0764 $\pm $0.0464  \\
$\Lambda$& 1.114  $\pm$0.047 & 0.0698 $\pm$ 0.0148  \\
$\Xi^-$  & 1.110  $\pm$0.218 & 0.0440 $\pm$ 0.0752  \\
\hline  
\end{tabular}
\caption{Fitted values of the $T$ and $q$ parameters for strange particles 
measured by
the ALICE~\cite{Aamodt:2011zj} and CMS collaborations~\cite{Khachatryan:2011tm} 
using the Tsallis-B form for the momentum distribution. 
}
\end{center}
\end{table}
The striking feature is that the values of $q$ are consistently between 1.1 and 1.2
for all species of hadrons.  The fit to negatively charged particles in \pp collisions measured by the ALICE collaboration is shown in Fig.~\ref{pikp}.
\section{Antimatter}
One of the striking features of particle production at high energies
is the near equal abundance of matter and antimatter in the central rapidity 
region~\cite{star:2009,Aamodt:2010d}. As is well known, a similar symmetry 
existed in the initial stage of the early universe and it still remains a
mystery as to how this got lost in the subsequent evolution of the
universe reaching a stage with no visible amounts of antimatter being present.

Closely related to this matter/antimatter symmetry is the production
of light antinuclei, hypernuclei, and antihypernuclei at high energies.
Since the first observation of hypernuclei in 1952 \cite{first_hypernucleus},  
there has been a steady interest in searching for new hypernuclei, exploring the hyperon-nucleon
interaction which is relevant (see, e.g., \cite{hahn,stoecker}) for nuclear physics.
Hypernuclei decay with a lifetime which depends on the strength of
the hyperon-nucleon interaction. While several hypernuclei have
been discovered since the first observations in 1952, no antihypernucleus has ever been
observed until the recent discovery of the antihypertriton in
Au+Au collisions at $\sqrt{s_{NN}}$ = 200 GeV by the STAR collaboration at
RHIC \cite{Abelev:2010}. 
The yield of (anti)hypernuclei  measured  by STAR is very large, in particular, they seem
to be produced with a  similar yield as other (anti)nuclei, in particular (anti)helium-3.
This  abundance is much higher than measured for hypernuclei and nuclei at 
lower energies~\cite{shuryak}.
It is of interest to understand the nature of this enhancement, and
for this the mechanism of production of (anti)hypernuclei should 
be investigated.

The thermalization assumption applies  successfully to hadrons
produced in a large number of particle and nuclear reactions 
at different energies (see, e.g., \cite{becattini1,review,energy}).
This fact allows us to estimate thermal parameters characterizing the particle 
source for each colliding system,  relevant for the understanding 
of the thermal properties of dense and hot matter, and in particular
for studies of QCD phase transitions. 
 
Using the parameterizations of thermal parameters found in the 
THERMUS model~\cite{thermus1,thermus2},  estimates have been made of the yields 
of (anti)hyper\-nuclei, that can be directly compared to 
the recently measured yields at RHIC,  
as well as of (anti)matter and (anti)hypernuclei production at 
the Large Hadron Collider (LHC)~\cite{sharma}.  
A similar analysis, not including 
\pp  results,  has been presented recently in~\cite{andronic-heavy} 
where it was shown that  ratios of hypernuclei to nuclei show an energy
dependence similar to the $K^+/\pi^+$ one with a clear maximum at lower energies.

A quantitative study as to how the matter/antimatter symmetry is reached as the
beam energy is increased has been presented in~\cite{sharma};
 estimates of ratios of hypernuclei and
antihypernuclei yields in Au+Au collisions at RHIC using
the above mentioned parameterizations of thermal
parameters that best fit hadron
production at RHIC have also been presented~\cite{sharma}.
The analysis uses a thermal model 
and  aims to  elucidate the production mechanism of hypernuclei and
 antihypernuclei in heavy ion collisions at RHIC and LHC energies,
thus providing insight in the surprising increase of (anti)hypernuclei production
at high energies.

\begin{figure}
\begin{center}
\includegraphics[width=0.8\linewidth,height=5cm]{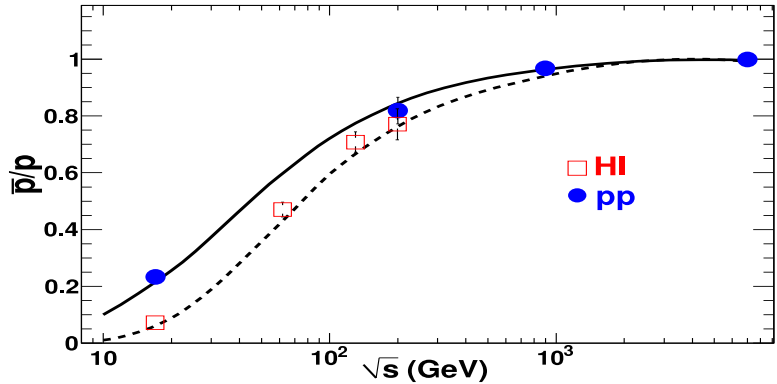}
\caption{The  \pbar/p ratio as  function of $\sqrt{s}$.
The solid circles are results from \pp collisions and the open squares
are results from HI collisions as a function
 of the invariant beam 
energy\cite{Abelev:2006cs,star:2009,Aamodt:2010d,Alt:2005gr,Alt:2007fe}.}
\label{pbarp}
\end{center}
\end{figure}

In heavy-ion collisions the increase in the  antimatter to matter ratio
with the center-of-mass energy of the system has been observed
earlier by the NA49~\cite{Alt:2005gr,Alt:2007fe} and the
STAR~\cite{Abelev:2006cs} collaborations.
The trend of the \pbar/p ratio increase with the energy towards unity
is shown in Fig.~\ref{pbarp}, where the open squares refer
to heavy ion collisions and the solid circles refer to \pp collisions.
It includes results from the NA49~\cite{Alt:2005gr}, STAR~\cite{Abelev:2006cs} 
and the new results from the ALICE collaboration~\cite{Aamodt:2010d}.
The two input parameters, the chemical freeze-out temperature $T$ and
the baryon chemical potential {\muB} as a function of $\sqrt{s}$ are
taken from Ref.~\cite{comparison}.
The solid circles represent {\muB}, obtained after fitting
experimental data with the THERMUS model~\cite{thermus1,thermus2}.
The solid line is a new parameterization adjusted for \pp 
collisions~\cite{sharma}. In view of the fact that peripheral and central 
collisions show no noticeable change in the temperature,  the
same $T$ dependence for \pp as in heavy ion collisions 
was used~\cite{sharma}. It is important to note that {\muB} is always lower
in \pp collisions than in heavy ion collisions,
e.g., the freeze-out chemical potential follows a different pattern, 
due to the lower stopping power in \pp collisions.

The relation between the \pbar/p ratio and {\muB} can be shown
easily within the statistical concept using the Boltzmann
statistics. 
The production of light nuclei including hypertritons
 ($^3_\Lambda$H) and antihypertritons ($^3_{\bar{\Lambda}}\overline{\textrm{H}}$)
was recently observed by the STAR collaboration~\cite{Abelev:2010}.
The abundances of such light nuclei and antinuclei follow a 
consistent pattern in the thermal model. 
The temperature remains the same as before but an extra 
factor of $\mu_B$ is picked up each time the baryon number is increased.
Each proton or neutron thus simply adds a factor of $\mu_B$ to the 
Boltzmann factor. 
The production of nuclear fragments is therefore very sensitive to the 
precise value of the baryon chemical potential and could thus lead
to a precise determination of $\mu_B$.
Deuterium has an additional neutron and the antideuterium to 
deuterium ratio is given by the square of the antiproton to 
proton ratio:
\begin{equation}
\frac {n_{\rm\overline{d}}}{n_{\rm d}}       =  e^{-(4 \mu_B )/T}.
\end{equation}

Helium-3 has 3 nucleons and the corresponding antihelium-3 to
helium-3 ratio is given by
\begin{equation}
\frac {n_{\rm ^3\overline{He}}}{n_{\rm ^3He}} =  e^{-(6 \mu_B)/T}.
\end{equation}
If the nucleus carries strangeness, this leads to an extra factor of $\mu_S$
\begin{equation}
\frac {n_{\rm ^3_{\overline{\Lambda}}\overline{H}}}{n_{\rm ^3_{\Lambda}H}} =
e^{-(6 \mu_B - 2 \mu_S )/T}.
\end{equation}
In mixed ratios,  the different degeneracy factors are also taken
into account, e.g., 6 for $^3_\Lambda H$ 
and 2 for $^3_\Lambda H$ 
\begin{equation}
\frac {n_{\rm ^3_{\Lambda}H}}{n_{\rm ^3He}}  = 3e^{-(6 \mu_B
- \mu_S )/T}.
\end{equation}
In the model like in the data the $He^3$ and $\overline{He^3}$ yields 
have been corrected  for the part coming from hypertriton and antihypertriton decays
assuming a decay branching ratio for the decay of 25 \%.
%
%
%
\section{Conclusions}
The thermal model is providing valuable insights in the composition of 
the final state produced in  
heavy ion collisions and also in \pp collisions.
It shows a clear systematic way of interpreting results 
concerning identified particles. The production of antimatter like antinuclei, hypernuclei and 
antihypernuclei  shows a new region of applications for the 
thermal model which promises to be very useful.
\section*{Acknowledgments} 
Numerous discussions with A. Kalweit, K. Redlich, H. Oeschler, N. Sharma, 
D. Worku, S. Kabana and I. Kraus are at the basis of the
results presented here. The financial support 
of the South Africa - Poland scientific collaborations is 
gratefully acknowledged.
%


\end{document}